\documentclass[12pt]{IOPART}
\usepackage{graphicx}
\newcommand{\gtsim}{\mbox{{\raisebox{-0.4ex}{$\stackrel{>}{{\scriptstyle\sim}}
$}}}}

\begin{document}
\title{On the de Haas-van Alphen effect in inhomogeneous alloys.}
\author{Neil Harrison$^1$ and John Singleton$^{1,2}$}
\address{$^1$National High Magnetic Field Laboratory, LANL,
MS-E536, Los Alamos, New Mexico 87545, USA}
\address{$^2$University of Oxford, Deaprtment of Physics, The
Clarendon Laboratory, Parks Road,
Oxford OX1 3PU, U.K.}


\begin{abstract}
We show that Landau level broadening in alloys occurs naturally as
a consequence of random variations in the local quasiparticle
density, without the need to consider a relaxation time.
This approach predicts
Lorentzian-broadened Landau
levels similar to those derived
by Dingle using the relaxation-time approximation.
However, rather than being determined by a finite relaxation time $\tau$,
the Landau level widths instead depend directly on the rate
at which the
de Haas-van Alphen frequency changes with alloy composition.
The results are in good agreement with recent data from
three very different alloy systems.
\end{abstract}

\submitto{\JPC}

\maketitle

The relaxation-time approximation (RTA), in which quasiparticles
scatter in random events at a characteristic
rate $\tau^{-1}$,
has proved invaluable in
understanding the electrical resistivity
and thermal conductivity of metals~\cite{ashcroft1}.
It has also been used to treat
the de~Haas-van~Alphen (dHvA) effect, {\it i.e.}
magnetic quantum oscillations
of the magnetisation~\cite{shoenberg1,lifshits}.
Dingle showed that the Lorentzian broadening of Landau levels in
metals, observed using the dHvA,
is qualitatively explained by the RTA for elastic scattering~\cite{dingle1}.
The predicted exponential damping of quantum
oscillations due to impurities, with a characteristic scaling temperature
known as the Dingle temperature $T_{\rm D}=\hbar/2\pi k_{\rm B}\tau$,
was subsequently verified experimentally in the vast majority of known
metals~\cite{shoenberg1,lifshits}.

In spite of this apparent
success, the RTA proves inadequate when considering
thermodynamic functions of state, such as the
electronic specific heat $\gamma T$ or magnetic susceptibility $\chi$,
because of the formation of bound states associated with
impurities~\cite{hewson1}.
More generally,
impurities, surfaces or defects of any kind will modify the lattice periodicity
and the local potential of the crystal, so that the quasiparticle wavefunctions
({\it i.e.} the eigenstates of a perfect, periodic, infinite crystal~\cite{ashcroft1})
will no longer be eigenstates of the modified crystal~\cite{ashcroft1}.
The assumption that the consequent alteration
of, for example, the Landau level structure can be parameterised
merely by a relaxation rate $\tau^{-1}$~\cite{dingle1}
seems to be a considerable
oversimplification.

A deeper problem arises in the alloy Hs$_{1-x}$Do$_x$,
where Hs represents the host material and Do the dopant
impurity.
When $x$ becomes $\sim 0.5$, one should be
concerned as to how far the distinction between ``host''
and ``dopant'' can be stretched.
As numerous experiments have shown (see {\it e.g}
References~\cite{springsteen,wasserman1,young,vuillemin1,badpoor1}),
the quasiparticle wavefunctions are no longer those
of the host material, weakly perturbed by enhanced
impurity scattering.
By contrast, Dingle's model only considers the statistical
broadening of Landau levels, without considering how the
underlying bandstructure is being modified by the alloying~\cite{dingle1}.

In view of the difficulty in separating effects due to
the variation of the bandstructure with $x$
and effects due to the changes in the actual quasiparticle lifetime,
we have chosen to abandon the RTA
and instead adopt a semi-empirical approach based on
statistical variations of the local quasiparticle density
in an alloy. The essential advantage of this approach is that the
sensitivity of the electronic structure to $x$ is considered as the
starting point.
Our model is
therefore related to the work of Woltjer~\cite{woltjer}.
Using a spatially-varying electron density,
Woltjer was able to provide convincing
simulations of Shubnikov-de Haas oscillations
and the Quantum Hall Effect in two-dimensional semiconductor
systems without the need to invoke localisation~\cite{woltjer,woltjerpuff}.
However, in contrast to Woltjer's work,
which involved numerical simulations of experimental data,
in the current paper we provide an analytical solution which
predicts the damping of dHvA oscillations.

Our model is applicable to systems for which the lattice parameters
are only weakly dependent on $x$~\cite{getout}
and for dHvA oscillations arising from circular Fermi surface cross-sections.
Encouragingly, it predicts
Lorentzian-broadened Landau
levels similar to those described by Dingle~\cite{dingle1}.
However, rather than being determined by a finite relaxation time $\tau$,
the Landau level widths instead depend directly on the rate
at which the
dHvA frequency $F(x)$ changes with $x$, enabling
estimates to be made that compare very favourably with
experiment.

The starting point of the model is the fact that $x$,
the local concentration of Do in Hs,
is subject to statistical
variations about the mean value $\bar{x}$.
This will lead to a spread of
Fermi-surface cross-sections $A(x)$ about the mean $A(\bar{x})$.
As the dHvA frequency is given by $F=(\hbar/2\pi e)A$~\cite{shoenberg1},
the variation in $A$ in turn leads to a spread of dHvA frequencies $F(x)$
and to phase smearing effects~\cite{shoenberg1,dingle1}.
In this paper we restrict ourselves to simple circular Fermi surface
cross-sections of $k$-space area $A$
that are easy to relate to the local
quasiparticle concentration $N(x)$~\cite{ashcroft1}.
However, rather than assuming that
$N(\bar{x})$ is linearly
dependent on $\bar{x}$, we choose a semi-empirical approach
whereby $A^\prime(\bar{x})$ and $F^\prime(\bar{x})$, the derivatives respectively
of the mean cross-section
$A(\bar{x})$ and mean dHvA frequency $F^(\bar{x})$
with respect to $\bar{x}$, are those obtained from
experiment.
The dHvA frequency $F(x)$ corresponding to
a particular value of $x$ can then be obtained from
\begin{equation}
F(x)\approx F(\bar{x})+[x-\bar{x}]F^\prime(\bar{x}).
\end{equation}

Given the finite separation, $a$, between ions~\cite{note1}, the
average number of Hs and Do ions encountered
in a quasiparticle path comprising $p$ orbits
of the circular cyclotron trajectory
is determined by binomial statistics;
the path will be of length
$2\pi pl_{\rm c}(\bar{x})=2\pi p\sqrt{2\hbar F(\bar{x})/eB^2}$.
For such a path, $m \approx xn$ of
these will be of the Do type and $n-m \approx [1-x]n$ of these will be of the
Hs type. The probability that $m$ of these ions are of type Do,
corresponding to a local dopant concentration $x=m/n$ (and local
dHvA frequency $F(x)$),
is therefore
\begin{equation}\label{prob}
    p(m,n)=\frac{\bar{x}^m[1-\bar{x}]^{n-m}n!}{m![n-m]!}.
\end{equation}
Under standard experimental conditions, $n$ will
always be a large number in metals. Taking the ``necks'' of Au as one
example of a Fermi surface cross section ($F_{\rm N}\approx$~1530~T)
that is not especially large~\cite{lifshits}, $a\sim$~2.6~\AA~
while $l_{\rm c}\sim$~1400~\AA~
in a magnetic field of $B\sim 10$~T, implying that
that $n\sim 540$. Clearly, it would be impractical to
work with such a large number terms in calculations.
It is well known, however,
that the skew factor $\eta=1/\sqrt{6\bar{x}[1-\bar{x}]n}$ for
the binomial distribution vanishes for large $n$, causing the
binomial distribution to become well approximated by the normal
distribution~\cite{guenault}. In implementing this approximation, the mean becomes
$\mu_m=n\bar{x}$ and the variance becomes
$\sigma^2_m=n\bar{x}[1-\bar{x}]$.
Following the established idea that a variation in
$F$ results in a ``phase-smearing'' which produces
a damping of the dHvA effect~\cite{shoenberg1,dingle1}
the damping factor becomes the result of the Fourier
transformation
\begin{eqnarray}\label{guassian}
    R_{\rm i}\approx\int_{-\infty}^{+\infty}\frac{n}{\sigma_m}
    \exp{\bigg(\frac{-[x-\bar{x}]^2n^2}{2\sigma_m^2}\bigg)}
    \nonumber\\
    \times\cos{\bigg(\frac{2\pi
    p[x-\bar{x}]F^\prime(\bar{x})}{B}\bigg)}{\rm d}x.
\end{eqnarray}
Making the substitution $\phi\equiv 2\pi p[x-\bar{x}]F^\prime(\bar{x})/B$,
and performing the integration in the $\phi$ domain, we obtain
\begin{eqnarray}\label{transform}
    R_{\rm i}\approx\int_{-\infty}^{+\infty}
    \frac{1}{\sigma_\phi}
    \exp{\bigg(\frac{-\phi^2}{2\sigma_\phi^2}\bigg)}
    \cos{(\phi)}{\rm d}\phi\nonumber\\
    \approx\exp{\bigg(\frac{-\sigma_\phi^2}{2}\bigg)},
\end{eqnarray}
where
\begin{equation}\label{phase}
    \sigma_\phi^2=\frac{2\pi p\bar{x}[1-\bar{x}]F^\prime(\bar{x})^2a}
    {B}\sqrt{\frac{e}{2\hbar F}}
\end{equation}
is now the phase-variance.

The expected symmetry between Hs and Do is immediately seen in
the presence of both $[1-\bar{x}]$ and $\bar{x}$ terms in
Equation (\ref{phase}), enabling this model to be applied across an
entire alloy series $0$~$<x<$~1. More satisfyingly, since the exponent
is linear in both the harmonic index $p$ and $1/B$, the functional
form of $R_{\rm i}$ is exactly that obtained by Dingle
using the RTA~\cite{dingle1}.
After inverse Fourier
transformation and the parabolic band substitution
$F=m^\ast E/\hbar e$, a Lorentzian Landau-level line shape
\begin{equation}\label{lorentzian}
    f(E)=\frac{\Gamma}{\pi[E^2+\Gamma^2]},
\end{equation}
is obtained, with the level width $\Gamma$ given by
\begin{equation}\label{width}
    \Gamma=\frac{\bar{x}[1-\bar{x}]F^\prime(\bar{x})^2a}{m^\ast}
    \sqrt{\frac{\hbar e^3}{8F}}.
\end{equation}
This latter result implies that a Lorentzian broadening of the Landau
levels results naturally from frequency-smearing effects caused by the
substitution of dopants without needing to consider the concept of
a relaxation time~\cite{woltjerpuff}.

We now consider whether this model can
account for a significant amount of the Landau-level broadening
observed in well known alloy systems. Three
experiments involving approximately circular
Fermi-surface cross-sections are considered.

\noindent
{\it Ag impurities in Au.}
Dilute alloys of the form
Ag$_x$Au$_{1-x}$ provide a useful test case, since the lattice
parameters of Au and Ag are very similar and the neck orbit, giving
rise to a dHvA frequency of $F_{\rm N}\approx$~1530~T in Au, is thought to
be very circular~\cite{lifshits}.
To facilitate a comparison with existing experimental
data, the results of our model are stated in terms of
an effective Dingle temperature
\begin{equation}\label{dingle}
    T_{\rm D}=\frac{\bar{x}[1-\bar{x}]F^\prime(\bar{x})^2a}
    {\pi k_{\rm B}m^\ast}\sqrt{\frac{\hbar e^3}{2F}}.
\end{equation}
On inserting the appropriate values for
Ag$_{0.01}$Au$_{0.99}$, of $F^\prime_{\rm N}\sim$~650~T,
$m^\ast\sim$~0.29~$\times m_{\rm e}$ and
$a\sim$~2.7~\AA~
for the face-centered cubic lattice~\cite{ashcroft1,shoenberg1,lifshits},
we obtain $T_{\rm D}\approx 1.1$~K per percent of Ag, which compares favourably with the
experimentally obtained value of 0.8~K per percent of
Ag~\cite{springsteen,vuillemin1}. Thus our
model for Landau-level broadening appears to be able to predict
reasonable values for the observed $T_{\rm D}$
in Ag$_x$Au$_{1-x}$.

\noindent
{\it Kondo alloys.}
Our model can also be applied to the
Ce$_x$La$_{1-x}$B$_6$ series. Because this is a Kondo system in which
the effective mass varies with the magnetic field, it is
more meaningful to present the results in terms of a
mean free path, since this quantity is not renormalised by the
interactions that give rise to Kondo behaviour~\cite{toyotaearly}.
In having abandoned the RTA, however, we have in effect also abandoned the
concept of a mean free path. Nevertheless,
by making a comparison with the formulae of Dingle,
we can define an {\it effective mean free path},
\begin{equation}\label{path}
    l_{\rm eff}=\frac{2\hbar F}{e\bar{x}[1-\bar{x}]F^\prime(\bar{x})^2a}.
\end{equation} 
On inserting the appropriate values of $F_{\alpha,3}\sim$~7970~T,
$F^\prime_{\alpha,3}\sim$~550~T and $a\sim$~4.0~\AA,~for the
worst-case alloy $x=$~0.5~\cite{badpoor1},
we obtain $l_{\rm eff}\approx~350$~nm, which is
within experimental error of the maximum value obtained
experimentally~\cite{badpoor1}.
Thus, this model can explain why the broadening of the Landau levels
in the Ce$_x$La$_{1-x}$B$_6$ intermetallic compounds was observed to be unexpectedly
low~\cite{badpoor1}.

\noindent
{\it A doped insulator.}
Finally, it is interesting to
consider the case of a doped insulator, to investigate whether
one should expect to observe the dHvA in such systems. Low density,
weakly-ferromagnetic electron-gas systems, which have been of recent
interest, certainly fall into the category, and dHvA oscillations have
been observed~\cite{young}. The simplest model is that of a variable
density, $N$, where $N(x)=xN^\prime$. On deriving $F(x)$
and $F^\prime(x)$ for a spherical Fermi surface, we obtain
\begin{equation}\label{insulator}
    \frac{-\sigma_\phi^2}{2}=\frac{p}{24}[1-x]
    \frac{\hbar a}{eB}N^\prime.
\end{equation}
Interestingly, at very low concentrations, $x\rightarrow$~0, the
extent to which the quantum oscillations are damped does not depend
on $x$. Since no estimates of the $T_{\rm D}$ or $l_{\rm eff}$
have been published \cite{young}, the best we can do is
estimate the lowest field at which we should expect quantum
oscillations to be observed. According to Dingle~\cite{dingle1}, the
threshold field is given approximately by the inequality
$\omega_{\rm c}\tau\gtsim 1$. The equivalent inequality according to our model is
$\phi^2\gtsim 2\pi$. Upon substituting $a\sim$~4.0~\AA~
and $N^\prime\sim$~5$\times$10$^{27}$~m$^{-3}$~\cite{young}, we obtain $B\gtsim$~9~T.
This is in very good agreement with experiment;
Goodrich~\cite{perscom} reports that dHvA oscillations are only observed at
fields of 10~T and greater.

In summary, we have shown that Lorenzian broadening of Landau levels
in alloys can be derived statistically by considering a
distribution of Fermi surface cross-sections, without
invoking the relaxation-time concept.
We have considered three alloy systems, and shown that the
extent to which the Landau levels are broadened, or the extent to
which the quantum oscillations are damped, compares
favourably with experiment.
In view of the current experimental interest in alloy systems,
we hope that this paper stimulates further
consideration of, for example, more general Fermi-surface shapes
and the effect of lattice mismatch between host and impurity.

Work at Los Alamos is supported by the Department of Energy, the National
Science Foundation (NSF) and the State of Florida.
JS acknowledges additional support by EPSRC (UK).
We should like to thank Charles Mielke and Albert Migliori for very
helpful discussions.


\begin{thebibliography}{9}
\bibitem{ashcroft1} N.~W.~Ashcroft and
N.~D.~Mermin, \textit{Solid State Physics} (Saunders College Publishing,
1976).
\bibitem{shoenberg1} D.~Shoenberg, \textit{Magnetic
Oscillations in Metals} (Cambridge University Press, Cambridge 1984).
\bibitem{lifshits}
I.M. Lifshits, M.Ya. Azbel and M.I. Kaganov,
{\it Electron theory of metals} (Consultants Bureau,
New York 1973).
\bibitem{dingle1}  R.~B.~Dingle, Proc. Roy. Soc. A {\bf 211}, 517
(1952).
\bibitem{hewson1} See, for example, A.~C.~Hewson, \textit{The Kondo
Problem to Heavy Fermions} (Cambridge University Press, Cambridge 1993).
\bibitem{springsteen} D.~H.~Lowndes, K.~M.~Miller, R.~G.~Poulsen and
M.~Springford, Proc. R. Soc. Lond. A. {\bf 331}, 497 (1973).
\bibitem{wasserman1} See A.~Wasserman and M.~Springford, Advan.
Phys. {\bf 45}, 471 (1996) and references therein.
\bibitem{young} D.~P.~Young, D.~Hall, M.~E.~Torelli, Z.~Fisk,
J.~L.~Sarrao, J.~D.~Thompson, H.~R.~Ott, S.~B.~Oseroff, R.~G.~Goodrich
and R.~Zysler, Nature {\bf 397}, 412 (1999).
\bibitem{vuillemin1} J.~J.~Vuillemin, preprint, to be published (2001).
\bibitem{badpoor1} R.~J.~Goodrich, N.~Harrison, A.~Teklu, D.~Young
and Z.~Fisk, Phys. Rev. Lett. {\bf 82}, 3669 (1999).
\bibitem{woltjer}
R. Woltjer {\it et al.}, Springer Series in Solid State Sciences,
(edited G. Landwehr), volume 71, page 104 (1986)
and volume 87, page 66 (1988).
\bibitem{woltjerpuff}
By analogy with the conclusions of the present paper,
Woltjer's work might also explain why Landau levels appear
to have Lorentzian line shapes
in two-dimensional systems~\cite{potts1}
for which models based on the RTA predict
otherwise~\cite{theory1,chakraborty1}.
\bibitem{potts1} A. Potts, R.~Shepherd, W.~G.~Herrendon-Harker,
M.~Elliott, C.~L.~Jones, A.~Usher, G.~A.~C.~Jones, D.~A.~Ritchie,
E.~H.~Linfield and M.~Grimshaw, J. Phys.: Condens. Matt. {\bf 8},
5189 (1996).
\bibitem{theory1}
T. Ando, B. Fowler and F. Stern,
Rev. Mod. Phys. {\bf 54}, 437 (1982).
\bibitem{chakraborty1} T.~Chakraborty and P.~Pietil\"{a}inen,
\textit{The Quantum Hall Effects} (Springer-Verlag, Berlin Heidelberg,
1995).
\bibitem{getout}
It remains to be tested whether
our approach can adequately accommodate large lattice mismatches
between Hs and Do.
\bibitem{guenault}
See {\it e.g.}
{\it Statistical Physics}, by Tony Guenault
(Routledge, London 1988)
\bibitem{note1} For a simple cubic system $a$ is the lattice
parameter, while more generally, $a=N^{\frac{1}{3}}$, where $N$ is
the total ion concentration inclusive of both Hs and Do.
\bibitem{toyotaearly}
N.~Toyota, E.W.~Fenton, T.~Sasaki
and M.~Tachiki, Solid State Commun.
{\bf 72}, 859 (1989).
\bibitem{perscom}
Roy Goodrich (2001), private communication.
\end{thebibliography}
\end{document}